\documentstyle[11pt,aaspp]{article}
\newcommand{\BE}{\begin{equation}}
\newcommand{\BEAL}{\begin{eqnarray}}
\newcommand{\EE}{\end{equation}}
\newcommand{\EEAL}{\end{eqnarray}}
\def\grad{\vec \nabla}
\def\rar{\rightarrow}
\def\FTO{~\_{2}F\_{1}}
\def\_#1{_{\scriptscriptstyle #1}}
\def\&#1{^{\scriptscriptstyle #1}}
\def\abs#1{\vert #1\vert}
\def\pd#1#2{{\partial#1\over\partial#2}}
\def\deriv#1#2{{d#1\over d#2}}
\def\RR{\Delta}
\def\l{\lambda}
\def\k{\kappa}
\def\o{\omega}
\def\n{\nu}

\def\a{\alpha}

\def\c{\gamma}

\def\rr{\varrho}
\def\ff{\varphi}
\def\z{\zeta}
\def\ro{R_{o}}
\def\Shat{\hat\Sigma}

\def\tssp{\tau\sin\&{2}\phi}

\def\Sia{\Sigma\&{\a}(r)}

\begin{document}
\title{Finite Disks with Power-Law Potentials}
\author{ Rafael Brada and Mordehai Milgrom}
\affil{Department of Condensed-Matter
 Physics, Weizmann Institute of Science  76100 Rehovot, Israel}
\begin{abstract}
 We describe a family of circular, and elliptical, {\it finite} disks
 with a disk potential that is a power of the radius. These are all
 flattenned ellipsoids, obtained by squashing finite spheres with a
 power-law density distribution, cutoff at some radius $\ro$. First we
 discuss circular disks whose circular rotation speed $v
 \propto r\&{\a}$, with any $\a> -1/2$. The surface-density of the
 disks is expressed in terms of hypergeometric functions of the radius:
 $ \Sia\propto (r/\ro)\&{2(\a-1)}[1-(r/\ro)\&{2}]\&{1/2}
  \FTO(1-\a,1/2;3/2;1-\ro\&{2}/r\&{2})$.
 The disk then has the same rotation curve as the sphere within
the cutoff radius (up to a constant factor). We give closed expressions
 for the full 3-D potentials in terms of hypergeometric functions of
 two variables. We express the potential and  acceleration
in the plane at $r>\ro$, and along the rotation axis, in terms of simple
 hypergeometric functions. All the multipoles of the disk are given.
We then generalize to non-axisymmetric disks with midplane axis ratio
 $p<1$. The potential in the midplane is given by $\varphi
(r,\phi)\propto r\&{2\a}F\_{1}(\a+1/2,-\a,\a+1/2,\a+1;\tssp;\tau)+c$,
 where $F\_{1}$ is the hypergeometric function of two variables, and
 $\tau=1-p\&{2}$. For integer values of $2\a$ the above quantities are
 given in more elementary terms. All these results follow
 straightforwardly from formulae we derive for the general, cutoff,
power-law, triaxial ellipsoid.
\end{abstract}
\section{Introduction}
\par
Models of disk galaxies with explicit expressions for the disk potential
 and for the surface density are quite useful in studies
of galactic structure, for example as initial states for
studies of stability, spiral-arm formation etc..
Infinite-extent, power-law disk models--with power-law rotation curves:
 $v(r)\propto r\&{\a}$, and power-law surface densities:
$\Sigma(r)\propto r\&{2\a-1}$--have their uses in this context
 (e.g \cite{BT87,EZ92,eva94}).
 They are, however, rather unrealistic.
\cite{qia92} describes an even wider class of
disks with known surface densities and potentials in the plane of the
disk; these too are of infinite extent.
We hope to contribute to the subject by describing a family of
 {\it finite} disks with power-law potentials (within the material
 disk). These are produced by squashing spheres with a power-law density
 profile, cutoff at some radius $\ro$, into ellipsoids (disks in the
 limiting case), the rotation curve in the equatorial plane, within
 $\ro$, is then the same as that of the sphere (up to a constant factor).
\par
\cite{mil89} used this method to determine the surface density of a
constant-acceleration (isodynamic) disk
with $v\propto r\&{1/2}$. We describe here the general case since,
as far as we know,
this useful class of models does not seem to be part of the known lore.
In the Appendix we derive expressions for the gravitational field of
a cutoff, power-law, triaxial ellipsoid. From these follow, after some
straightforward manipulations, all the results of the paper.
In section 2 we discuss thin, and thick axisymmetric disks;
in section 3 we derive their full 3-D potentials; and in section 4
we describe the generalization to non-axisymmetric disks, which
may also model galactic bars.
\section{Axisymmetric disk models}
\par
Start with a spherical mass distribution, with density
\BE \rho=AR\&{\c}, \label{i}\EE
 cutoff at $R=\ro$. Now squash the configuration in one direction,
by a constant factor $q<1$, to obtain an ellipsoidal mass distribution.
The gravitational acceleration, $g(r)$, at a point in the midplane,
 a distance $r\le\ro$ from the centre, in the equatorial plane,
 is determined only by the mass interior
to the ellipsoid going through $r$ (by ``Newton's third theorem''; see
e.g. \cite{BT87}). In particular, $g(r)$ is {\it independent
of the cutoff radius} $\ro$ (provided $\ro\ge r$).
The only dimensional parameter characterizing the system, on which
$g(r)$ may depend is then $A$, to which, in fact, $g$ must be
 proportional. On dimensional grounds, the only form that $g(r)$ can have
is then
\BE g(r)=-QAGr\&{\c+1}, \label{ii}\EE
where $Q$ is a dimensionless constant that depends on the axis ratio and
on $\c$. This argument generalizes to non-axisymmetric disks, and indeed
to the interior field of a cutoff, power-law, triaxial ellipsoid
(see section 4 and the Appendix).
We see that the equatorial rotation curves of this family of ellipsoidal
mass distributions is a power-law within the cutoff radius, with
 $v(r)=(QAG)\&{1/2}r\&{\a}$, $\a\equiv1+\c/2$.
Beyond $\ro$, $g(r)$ does depend on $\ro$, and the above dimensional
argument breaks down, as the expression for $g(r)$ may now contain
a factor $f(r/\ro)$.
\par
The surface density is just that of the spherical distribution
\BE \Sia=2A\int\_{0}\&{(\ro\&{2}-r\&{2})\&{1/2}}(r\&{2}+x\&{2})\&{\c/2}
{}~dx. \label{iii}\EE
Defining $u\equiv r/\ro$, $s\equiv[(\ro/r)\&{2}-1]\&{1/2}=
(u\&{-2}-1)\&{1/2}$, and $\Shat\equiv 2A\ro\&{\c+1}$,
 we can write after a change of variable
\BE \Sia={1\over 2}\Shat u\&{\c+1}s\int\_{0}\&{1}\l\&{-1/2}
(1+s\&{2}\l)\&{\c/2}~d\l.  \label{iv}\EE
This, in turn, can be written in terms of the hypergeometric function
\BE \Sia=\Shat u\&{2(\a-1)}(1-u\&{2})\&{1/2}\FTO(1-\a,1/2;3/2;-s\&{2}),
\label{v}\EE
(see \cite{GR80} 3.197.3).
We see that near the disk's edge, where $s\approx 0$, and
 $\FTO\approx 1$, $\Sia\approx\Shat(1-u\&{2})\&{1/2}$, for all the disks.
\par
  For integer values of $2\a$ the integrals in eqs.(\ref{iii})(\ref{iv})
 can be
 expressed in terms of elementary functions.
Some interesting special cases are: $\a=1$--which in the thin-disk limit
gives the Kalnajs disk (\cite{kal72})--has
\BE \Sigma(r)=\Shat(1-u\&{2})\&{1/2};  \label{spi}\EE
$\a=0$--which is the Mestel disk (\cite{mes63}) in the same limit--has
\BE \Sigma(r)=\Shat u\&{-1}\tan\&{-1}(u\&{-2}-1)\&{1/2}; \label{spii}\EE
 and the
isodynamic disk (\cite{mil89}) with constant acceleration ($\a=1/2$)
has
 \BE \Sigma(r)=\Shat \sinh\&{-1}(u\&{-2}-1)\&{1/2}. \label{spiii}\EE
The limit $\a\rar -1/2$ corresponds to a point mass at the origin (a
Keplerian rotation curve), and smaller values of $\a$ are excluded
as they correspond to an infinite total mass.
\par
  For $\a$ a positive integer, $\FTO$ in eq.(\ref{v}) is a polynomial of
order $\a-1$ in $s\&{2}$, and so, $\Sigma(r)$ is, up to the factor
$(1-u\&{2})\&{1/2}$, a polynomial of (even) order $2(\a-1)$ in $r$.
\par
Values of $\a>1$ correspond to disks with non-monotonic surface
 densities: for $1<\a<3/2$ $\Sigma(r)-\Sigma(0)\propto r\&{2(\a-1)}$
 near the origin; for $\a=3/2$ $\Sigma(r)-\Sigma(0)\propto-r\&{2}\ln r$
 there; and for $\a>3/2$ $\Sigma(r)-\Sigma(0)\propto r\&{2}$. For all
these cases $\Sigma$ is then increasing near the origin, but decreases
further out.
\par
The proportionality constant, $Q$, in expression(\ref{ii}) for $g$
 may be obtained by specializing results from Appendix A.
\BE Q(\k,\a)={4\pi\over 2\a+1}\FTO(1/2,\a+1/2;\a+3/2;\k),\label{fgds}\EE
where $\k\equiv 1-q\&{2}$.
\par
The interior potential for
 $R\le\ro[1+(\k/q\&{2})\cos\&{2}\theta]\&{-1/2}$
 (for $\c\not=-2$)
 as a function of the polar coordinates $R$ and $\theta$ is
\BE \varphi(R,\theta)={2\pi AG\over\a(2\a+1)}R\&{2\a+1}
  F\_{1}(\a+1/2,-\a,\a+1/2,\a+3/2;\k\sin\&{2}\theta,\k)+c,  \label{hkj}\EE
Where $F\_{1}$ is the first Appell function, or the
first hypergeometric
function of two variables (see \cite{GR80} 3.211), and
$c=-2\pi\a\&{-1}AG\ro\&{2\a}\k\&{-1/2}\sin\&{-1}\k\&{1/2}$.
\par
Hereafter we concentrate on infinitely thin disks, for which the results
may be put in simpler terms. For these
\BE Q=Q\_{0}(\a)=2\pi\&{3/2}\Gamma(\a+1/2)/\Gamma(\a+1).  \label{vi}\EE
\par
The potential in the disk is given (except for the Mestel case)
by
\BE \varphi(r)={Q\_{0}AGr\&{2\a}\over 2\a}+c.  \label{ffi}\EE
The constant $c$ can be determined from the potential outside
the disk treated in section 3:
\BE c=-{\pi\&{2}AG\ro\&{2\a}\over \a}=-{MG\over\ro}{\pi(\a+1/2)
\over 2\a}, \label{ffii}\EE
where
\BE M=4\pi A\ro\&{(2\a+1)}/(2\a+1) \label{relm}\EE
 is the total mass of the disk.
Note that $\varphi$, unlike $g$, does depend on $\ro$, in general.
\par
The total energy on a circular orbit at radius $r\le\ro$ may now be
 written
\BE E(r)=-{MG\over 2\ro}{\pi(\a+1/2)\over \a}\left[1-
\left({r\over\ro}\right)\&{2\a}
{(\a+1)\Gamma
(\a+1/2)\over \Gamma(1/2)\Gamma(\a+1)}\right].  \label{ffti}\EE
This energy is negative everywhere in the disk for $\a<1$,
 vanishes just at the edge, for the Kalnajs case,
and is positive in some region near the edge for $\a>1$, viz.
for $r>\eta(\a)\ro$, where
$\eta(\a)=[(\a+1)\Gamma(\a+1/2)/\pi\&{1/2}\Gamma(\a+1)]\&{-1/2\a}$.
It can be shown that $\eta<1$ for all $\a>1$, and we find numerically
that it has a minimum of $\approx 0.96$, at $\a\approx 4.37$; it is
thus only in a narrow region near the edge that $E(r)>0$.
We have seen that $\a>1$ disks have nonmonotonic surface densities.
Such disks may not be so realistic models by themselves;
they are useful as components when we expand the potential of
more general disks. We may find advantage, e.g., in  disks whose
 potential is a finite sum of power laws, some possibly with  $\a>1$.
\par
As an example, consider two of our disks with $\a\_{1}<\a\_{2}$,
 but with the same
$\ro$, and the same $\hat\Sigma=2A\ro\&{2\a-1}$ (the values of $A$ are
 different). The corresponding, unsquashed spheres then have the same
density at $\ro$.
 We see from expression(\ref{iii}) for the surface density
that, for a fixed $\hat\Sigma$, $\Sia$ is a monotonically decreasing
function of $\a$ for all $r<\ro$. Thus
\BE \Sigma\&{\a\_{1}\a\_{2}}\equiv \Sigma\&{\a\_{1}}-\Sigma\&{\a\_{2}}
\label{hjiy}\EE
is positive everywhere. The acceleration in such a disk is
\BE g=-\hat\Sigma G\pi\&{3/2}
\left[{\Gamma(\a\_{1}+1/2)\over\Gamma(\a\_{1}+1)}u\&{2\a\_{1}-1}
-{\Gamma(\a\_{2}+1/2)\over\Gamma(\a\_{2}+1)}u\&{2\a\_{2}-1}\right] \EE
($u=r/\ro$), and is directed inward everywhere, because
 $u\&{2\a-1}\Gamma(\a+1/2)/\Gamma(\a+1)$ is decreasing with $\a$.
 Furthermore, $\Sigma\&{\a\_{1}\a\_{2}}(r)$ vanishes at the edge
faster than $(1-u\&{2})\&{1/2}$; in fact, there
\BE \Sigma\&{\a\_{1}\a\_{2}}(r)\approx
 {2\over 3}\hat\Sigma(\a\_{2}-\a\_{1})(1-u\&{2})\&{3/2}, \EE
which can be seen by taking $\FTO$ to first order in $s\&{2}$ in
 eq.(\ref{v}).
We can get disks that vanish even faster at the edge by taking
the difference of two such disks with different pairs of $\a$'s
with the same difference. In the limit $\a\_{2}\rightarrow\a\_{1}$
we get a disk whose surface density is the derivative of $\Sia$
with respect to $\a$ {\it at a fixed $\hat\Sigma$}.

\section{The field outside the axisymmetric thin disk}
Building on formulae in Appendix A we get for the potential at
a general position outside the disk (for $\c\not=-2$--the Mestel
 disk requires separate treatment)
\BE \ff(r,z)=-{MG\over \ro}\hat\ff(r/\ro,z/\ro), \label{ti}\EE
 where the dimensionless
 potential $\hat\ff$, as a function of the dimensionless variables
$\rr\equiv r/\ro,~\z\equiv z/\ro$, is
\BE \hat\ff(\rr,\z)={\a+1/2\over \a}\left[\sin\&{-1}(1+L)\&{-1/2}
 -{1\over 2}\int\_{0}\&{W}{t\&{\a-1/2}(\RR\&{2}-\z\&{2}t)\&{\a}\over
(1-t)\&{\a-1/2}}~dt\right]. \label{ssiii}\EE
Here, $\RR$ is the dimensionless
 polar radius $\RR\&{2}\equiv\z\&{2}+\rr\&{2}$,
\BE L={1\over 2}\{\RR\&{2}-1+[(\RR\&{2}-1)\&{2}+4\z\&{2}]\&{1/2}\},
\label{ssiv}\EE
($L$ is $\o(1,\vec R)$ of appendix A, with $Y=\ro$), and
$W\equiv 1/(1+L)$.
The integral in eq.(\ref{ssiii}) may be identified, after some
 manipulations, as an integral representation of the first Appell
 function.  We thus get finally
\BEAL \hat\ff(\rr,\z)&=&{\a+1/2\over \a}\sin\&{-1}(1+L)\&{-1/2}-\nonumber
\\ &&-{1\over 2\a}\RR\&{2\a}W\&{\a+1/2}
  F\_{1}(\a+1/2,-\a,\a+1/2,\a+3/2;U,W),  \label{ssv}\EEAL
where
\BE U\equiv 1-L/\RR\&{2}.  \label{ssvi}\EE
\par
  For all non-zero integer values of $2\a$ the
 integral in eq.(\ref{ssiii}) can be performed and expressed in terms of
elementary functions.
 One case in point is the Kalnajs disk
which is just an homogeneous ellipsoid, for which the potential is given
in \cite{mac58} or in \cite{BT87}.
 Another instance is that of the constant-acceleration disk
where, integrating explicitly, we get for the dimensionless 3-D potential
\BE \hat\ff(\rr,\z)=2L\&{1/2}-2\RR
+2\sin\&{-1}(1+L)\&{-1/2}+\abs{\z}
\ln\left\{{(L\&{1/2}-\abs{\z})(\RR+\abs{\z})\over
(L\&{1/2}+\abs{\z})(\RR-\abs{\z})}\right\}.  \label{ri}\EE
 For the Mestel case with $\a=0$, eqs.(\ref{ssiii})
and (\ref{ssv}) give $0/0$, and it requires special treatment, which
 does not seem to yield a simple closed expression.
In general, all expressions for the accelerations hold for the $\a=0$
 case; those for potentials and energies require either a separate
derivation, or the application of
 the L'Hospital rule.
 For example, we can use this rule
with eq.(\ref{ssv}) to obtain for the Mestel case
\BEAL \hat\ff(\rr,\z)&=&-{1\over 2}(\ln\RR\&{2}+\ln W)
\sin\&{-1}(1+L)\&{-1/2}-\nonumber\\
 &&-{1\over 2}W\&{1/2}\pd{}{\a}F\_{1}(\a+1/2,-\a,\a+1/2,\a+3/2;U,W)
\bigr\vert\_{\a=0}. \label{ssttv}\EEAL
\par
The integral in eq.(\ref{ssiii}), for $\a\not=0$,
 can also be expressed in terms of simpler functions
when we confine ourselves to the plane ($z=0,~r>\ro$), or to the symmetry
axis ($r=0$). We then get for the dimensionless potential for the former
case
\BE \hat\ff(\rr>1,0)={2\a+1\over 2\a}\sin\&{-1}\rr\&{-1}
-{1\over 2\a\rr}\FTO(1/2,\a+1/2;\a+3/2;\rr\&{-2});  \label{rii}\EE
along the axis we have
\BE \hat\ff(0,\z)={2\a+1\over 2\a}\sin\&{-1}(1+\z\&{2})\&{-1/2}
-{1\over 2\a\abs{\z}}
\FTO(1,\a+1/2;\a+3/2;-\z\&{-2}).  \label{riii}\EE
[Equations(\ref{rii})(\ref{riii}) can also be obtained as special cases
 of eq.(\ref{ssv});
for $\z=0$ we have $U=W$, and for $\rr=0$ we have $U=0$; in both cases
$F\_{1}$ can be expressed as a hypergeometric function of one variable.]
\par
The acceleration in the plane can be brought to the form (valid also for
$\a=0$)
\BE g(r\ge\ro,z=0)=-{MG\over r\&{2}}
\FTO(1/2,\a+1/2;\a+3/2;\ro\&{2}/r\&{2}).  \label{viii}\EE
  For $r\rar\infty$ $\FTO\rar 1$, and we get the Keplerian expression.
 At the disk's edge
  $\FTO(1/2,\a+1/2;\a+3/2;1)=
\pi\&{1/2}(\a+1/2)\Gamma(\a+1/2)/\Gamma(\a+1)$
is the ratio of the disk's field to that of a sphere with the same mass
\par
The acceleration along the symmetry axis
 can be written as
\BE g(r=0,z)=-{MG\over z\&{2}}\FTO(1,\a+1/2;\a+3/2;-\ro\&{2}/z\&{2}).
  \label{xvii}\EE
\par
Using relations found in \cite{MO49}
 to get explicit expressions for $\FTO$ in special cases
we obtain from eqs.(\ref{viii}) and (\ref{xvii})
 (with $\hat t\equiv\rr\&{-1}=\ro/r$, and $t\equiv\z\&{-1}=\ro/z$):

 for the Kalnajs disk
\BE g(r\ge\ro,z=0)=-{MG\over r\&{2}}{3[\hat t~\sin\&{-1}\hat t-\hat t
\&{2}(1-\hat t\&{2})\&{1/2}]\over 2\hat t\&{4}} \label{ix}\EE
(reproducing a result of \cite{mac58}), and
\BE g(r=0,z)=-{MG\over z\&{2}}{3\over t\&{2}}\left(1-{\tan\&{-1}t\over t}
\right).   \label{xviii}\EE
  For the isodynamic disk
\BE g(r\ge\ro,z=0)=-{MG\over r\&{2}}{2\over 1+(1-\hat t\&{2})\&{1/2}}  ,
 \label{x}\EE
and
\BE g(r=0,z)=
-{MG\over z\&{2}}{\ln(1+t\&{2})\over t\&{2}}.  \label{xix}\EE
 For the Mestel disk we get the \cite{mes63} result for the plane
\BE g(r\ge\ro,z=0)=-{MG\over r\&{2}}\hat t
\&{-1}\sin\&{-1}\hat t, \label{xia}\EE
and
\BE g(r=0,z)=-{MG\over z\&{2}}{\tan\&{-1}t \over t},  \label{xxi}\EE
along the axis.
\par
Because $\FTO$ is a power series with simple expressions for the
 coefficients, any of eqs.(\ref{rii})(\ref{riii})(\ref{viii})(\ref{xvii})
 gives straightforwardly the multipoles of the disk:
 Write the potential (using polar coordinates) as
\BE \ff(R,\theta,\phi)=\sum\_{\ell=0}\&{\infty}\mu\_{\ell}
R\&{-(\ell+1)}P\_{\ell}(\cos\theta)  \label{xii}\EE
(only even $\ell$'s appear). Use, e.g., eq.(\ref{viii});
the hypergeometric function appearing in it is a power series
\BE \FTO(1/2,\a+1/2;\a+3/2;\ro\&{2}/r\&{2})=\sum\_{k=0}\&{\infty}f\_{k}
\ro\&{2k}r\&{-2k},   \label{xiv}\EE
with
\BE f\_{k}={\Gamma(k+1/2)(\a+1/2)\over \pi\&{1/2}k!(\a+k+1/2)}.
 \label{xv}\EE
The radius of convergence of this series in $\ro/r$ is 1 (i.e. the
expansion is valid for $r>\ro$).
Comparing with the $R$ derivative of expression(\ref{xii}), putting
$P\_{2k}(0)=(-1)\&{k}(2k)!2\&{-2k}(k!)\&{-2}$, and using some identities
for the $\Gamma$ functions, we get for the
multipole coefficients of the disk
\BE \mu\_{2k}=(-1)\&{k}GM\ro\&{2k}{\a+1/2\over (2k+1)(\a+k+1/2)}.
  \label{xvi}\EE
We see that, up to a numerical factor, $\mu\_{2k}$ is given by
 $GM\ro\&{2k}$, which is to be expected on dimensional grounds
as, in constructing $\mu$, we can only avail ourselves of
 the dimensional quantities $M$ and $\ro$ (or $A$ and
$\ro$).
  For example, the quadrupole moment is $\mu\_{2}=-GM\ro\&{2}(\a+1/2)/
3(\a+3/2)$. We reiterate that while expressions (\ref{ssiii})(\ref{ssv})
 for the potential are valid everywhere outside the disk, the
 multipole expansion is valid only outside the sphere
of radius $\ro$.
\section{Non-axisymmetric thin disk models}
\par
Our disks may be generalized to non-axisymmetric ones, which may
serve as models for non-axisymmetric galactic disks or for bars.
Start again with our cut-off-power-law sphere and squash it
by a factor $p$ along the $x$ axis and into a thin disk along the $z$
axis, to obtain a finite, elliptical disk of axis ratio $p$.
The field vanishes within a triaxial homoeoid obtained by squashing
a spherical shell of constant surface density by different factors
along different axes. Thus, our dimensional argument of section 1 still
holds, and the acceleration within the disk itself must still
be proportional to $r\&{\c+1}$, and oblivious to the cutoff.
Thus
\BE g\_{r}(r,\phi)=-AGr\&{\c+1}P\_{r}(\tau,\c,\sin\&{2}\phi),
 \label{IVi}\EE
\BE g\_{\phi}(r,\phi)=-AGr\&{\c+1}P\_{\phi}(\tau,\c,\sin\&{2}\phi),
 \label{IVii}\EE
where $r,\phi$ are polar coordinates,
and $\tau\equiv 1-p\&{2}$ is a measure of the departure from
 axisymmetry.
When $\c\not=-2$ eq.(\ref{IVi}) tells us that the midplane
 potential inside the disk must be of the form
\BE \varphi(r,\phi)={AGr\&{\c+2}\over \c+2}P\_{r}(\tau,\c,\sin\&{2}\phi)
+C(\phi,\ro)  \label{IViii}\EE
(note that the potential, unlike the acceleration, does depend on $\ro$).
Deriving $g\_{\phi}=-r\&{-1}\pd{\varphi}{\phi}$ from eq.(\ref{IViii})
and comparing with
 eq.(\ref{IVii}) we see that $C$ may not depend on $\phi$ and that
\BE P\_{\phi}={1\over \c+2}\pd{P\_{r}}{\phi}.  \label{IViv}\EE
All that remains is to derive one angular factor $P\_{r}$.
  For the special case $\c=-2$ eq.(\ref{IVi}) tells us that the potential
must be of the form
\BE \varphi(r,\phi)=AG\ln rP\_{r}(\tau,\c,\sin\&{2}\phi)
+C(\phi,\ro), \label{IVv}\EE
but now for eq.(\ref{IVii}) to hold we must have $\pd{P\_{r}}{\phi}=0$,
 $C=AG\hat C(\phi)+c(\ro)$, and $P\_{\phi}=\deriv{\hat C}{\phi}$.
We learn then that for this case the radial acceleration is
$\phi$-independent.
\par
The surface-density of the non-axisymmetric disk is obtained simply from
expression(\ref{v}) for the axisymmetric case by multiplying the latter
 by a factor $1/p$--to account for the squashing along $x$--and by
 replacing the variable $r$ in eq.(\ref{v})
 by $r[1+(\tau/p\&{2}) \cos\&{2}\phi]\&{1/2}$;
the boundary is at $r=\ro/[1+(\tau/p\&{2}) \cos\&{2}\phi]\&{1/2}$.
\par
We have derived the field inside the disk by integrating the
 contributions of thin triaxial homoeoids to the exterior potential
(taken from table 2-1 of \cite{BT87})
weighted by the power-law density profile. The derivation is given
in Appendix A.
The components of the acceleration may be written as
\BE g\_{r}=-2\pi AGr\&{2\a-1}\int\_{0}\&{1}
{t\&{\a-1/2}(1-\tssp~t)\&{\a}\over
(1-t)\&{1/2}(1-\tau t)\&{\a+1/2}}~dt,  \label{IVvi}\EE

\BE g\_{\phi}=2\pi AGr\&{2\a-1}\tau\sin\phi~\cos\phi
\int\_{0}\&{1}
{t\&{\a+1/2}(1-\tssp~t)\&{\a-1}\over
(1-t)\&{1/2}(1-\tau t)\&{\a+1/2}}~dt,  \label{IVvii}\EE
where $\a=1+\c/2$.
The integrals in eqs.(\ref{IVvi})(\ref{IVvii}) may be written in terms of
the first Appell function
\BE g\_{r}=-Q\_{0}AGr\&{2\a-1}F\_{1}(\a+1/2,-\a,\a+1/2,\a+1;
\tssp;\tau), \label{IVviii}\EE
\BEAL g\_{\phi}&=&Q\_{0}AGr\&{2\a-1}{(\a+1/2)\over (\a+1)}
\tau\sin\phi~\cos\phi\times \nonumber\\
 &&           \times F\_{1}(\a+3/2,-\a+1,\a+1/2,\a+2;\tssp;\tau).
   \label{IVix}\EEAL
These are valid for all values of $\c>-3$, including $\c=-2$ ($\a=0$).
In the last case we see from eq.(\ref{IVvi}) that indeed $g\_r$
does not depend on $\phi$.
\par
When $\c\not= -2$, the potential in the disk is given by
\BE \varphi(r,\phi)={Q\_{0}AGr\&{2\a}\over 2\a}
  F\_{1}(\a+1/2,-\a,\a+1/2,\a+1;\tssp;\tau)+c,   \label{IVviiii}\EE
where
\BE c=-{\pi\&{2}AG\ro\&{2\a}\over \a}\FTO(1/2,1/2;1;\tau)
\label{IVxi}\EE
is calculated in Appendix A. In the axisymmetric case
$\FTO=F\_{1}=1$ as the arguments vanish
 and we get the results of section 2.
\par
  For $\c=-2$ the interior potential is
\BEAL \varphi(r,\phi)&=&2\pi\&{2}AG
\FTO(1/2,1/2;1;\tau)\ln(r/\ro)+\nonumber
\\&   &+\pi AG\int\_{0}\&{1}{\ln[(1-\tssp
{}~t)/t]\over t\&{1/2}(1-t)\&{1/2}(1-\tau t)\&{1/2}}~dt-\nonumber\\
 &         &-\pi AG\int\_{0}\&{1}{\ln[(1-\tau t)/t\&{2})]
\over t\&{1/2}(1-t)\&{1/2}(1-\tau t)\&{1/2}}~dt,  \label{IVxii}\EEAL
to be compared with eq.(\ref{IVv}). The first term gives $g\_{r}$, the
second $g\_{\phi}$, and the third contributes to the
 constant that corresponds to
$\varphi=0$ at infinity.
\par
We next discuss various special cases.
As in the axisymmetric case non-zero integer values of $2\a$ afford
 simplifications. We get for $\a=1$:
\BE \varphi(r,\phi)={1\over 2}\pi\&{2}AGr\&{2}[\FTO(3/2,3/2;2;\tau)
-{3\tssp\over 4}\FTO(3/2,5/2;3;\tau)]+c,  \label{ccbt}\EE
which is simply an anisotropic, harmonic potential (known to be the case
inside an homogeneous ellipsoid--it is the basis, e.g., of the Freeman
bar model described by \cite{BT87}).
\par
  For the $\a=1/2$ disk we have
\BE \varphi(r,\phi)=4\pi AGrB(\tau,\sin\&{2}\phi)+c,  \label{mmi}\EE
where $c$ is given by eq.(\ref{IVxi}), and
\BE B={\abs{\cos\phi}\over (1-\tau)\&{1/2}\tau\&{1/2}}\tan\&{-1}
\left({\tau\cos\&{2}\phi\over 1-\tau}\right)\&{1/2}+
{\abs{\sin\phi}\over 2\tau\&{1/2}}\ln{1+\tau\&{1/2}\abs{\sin\phi}
\over 1-\tau\&{1/2}\abs{\sin\phi}}. \label{fffii}\EE
\par
The weak-asymmetry limit, $\tau\ll 1$, is readily obtained for
the a general $\a$, as $F\_{1}$
is a known power expansion in its two arguments. To first order in
$\tau$ we have
\BE \varphi(r,\phi)
={Q\_{0}AGr\&{2\a}\over 2\a}[1+{\tau(\a+1/2)\over
 2(\a+1)}(1+2\a\cos\&{2}\phi)]
-{\pi\&{2}AG\ro\&{2\a}\over \a}[1+{\tau\over 4}], \label{IVxiv}\EE
\BE g\_{r}=-Q\_{0}AGr\&{2\a-1}[1+{\tau(\a+1/2)\over
2(\a+1)}(1+2\a\cos\&{2}\phi)],
\label{IVxiii}\EE
and
\BE g\_{\phi}=Q\_{0}AGr\&{2\a-1}{(\a+1/2)\over (\a+1)}
\tau\sin\phi~\cos\phi.  \label{IVixaaa}\EE
\par
The disk potential and (radial) acceleration along the principal axes may
also be written in simpler terms for general $\a$.
Along the (shorter) $x$-axis, where $\phi=0$
\BE g\_{r}=-Q\_{0}AGr\&{2\a-1}\FTO(\a+1/2,\a+1/2;\a+1;\tau),
 \label{bnb}\EE
and along the longer axis
\BE g\_{r}=-Q\_{0}AGr\&{2\a-1}\FTO(1/2,\a+1/2;\a+1;\tau). \label{bkb}\EE
\par
The most general case pertains to the external field of a thick disk.
This is given in Appendix A
in terms of a Lauricella function of four variables,
$F\&{(4)}\_{D}$,
 which is a generalization of the hypergeometric function (see
 \cite{ext76}). In any of the three symmetry planes of the ellipsoid this
reduces to a Lauricella function of three variables $F\&{(3)}\_{D}$
($F\&{2}\_{D}$ is the same as $F\_{1}$).
 It is
 straightforward to obtain all these from formulae in Appendix A. All
these functions are single integrals between 0 and 1 of simple
algebraic functions. \cite{ext76} gives a short FORTRAN program
to evaluate these functions.
\acknowledgments

We thank David Earn for useful comments.
\appendix
\section{Formulae for the field of a cutoff, power-law ellipsoid}
\par
We start with the expression, given in Table 2-1 of \cite{BT87}
 for the potential at position $\vec R$
 outside a thin, triaxial homoeoid
of axes $a\_{i},~1=1,2,3$
\BE \varphi(\vec R)=-{G\over 2}M\_{shell}K(\vec b), \label{Ai}\EE
where $b\_{i}$ are the axes of the ellipsoid that is confocal with
the shell and passes through $\vec R$: $b\&{2}\_{i}=a\&{2}\_{i}+
\l(\vec a,\vec R)$, and $\l$ is defined by
\BE \sum R\&{2}\_{i}/ (a\&{2}\_{i}+\l)=1.  \label{Ao}\EE
 The quantity $K$ is given by
\BE K(\vec b)=\int\_{0}\&{\infty}{ds\over
 [(b\&{2}\_{1}+s)(b\&{2}\_{2}+s)
(b\&{2}\_{3}+s)]\&{1/2}}.  \label{Aii}\EE
We now calculate the potential outside a cutoff,
 power-law ellipsoid, whose external axes are $pY$, $Y$, and $qY$,
integrating the contributions of thin-shells of axes $kpY,~kY,~kqY$,
with $0\le k\le 1$.
The mass of the shell between $k$ and $k+dk$ is that of the original
spherical shell $dM=4\pi AY\&{\c+3}k\&{\c+2}dk$, so that
\BE \varphi(\vec R)=-2\pi AGY\&{\c+2}J, \label{Aiil}\EE
with
\BE J=Y\int\_{0}\&{1}k\&{\c+2}
dk\int\_{0}\&{\infty}
{ds\over \{[(pkY)\&{2}+\l+s][(kY)\&{2}+\l+s]
[(qkY)\&{2}+\l+s]\}\&{1/2}}.  \label{Aiii}\EE
Changing variables to $t\equiv (\l+s)/(kY)\&{2}$ we have
\BE J\equiv\int\_{0}\&{1}k\&{\c+1}
dk\int\_{\o(k,\vec R)}\&{\infty}
f(t)~dt,  \label{Aiv}\EE
with
\BE f(t)\equiv [(p\&{2}+t)(1+t)(q\&{2}+t)]\&{-1/2}  \label{Av}\EE
that is independent of $\vec R$, $k$, or $Y$, the dependence on which
 enters
 through the lower integration boundary $\o\equiv\l(k,\vec R)/(kY)\&{2}$.
We now integrate by parts over $k$ (the case $\c=-2$ requires a special
 treatment) to get
\BE J={1\over \c+2}\int\_{\o(1,\vec R)}\&{\infty}~f(t)~dt
+{1\over\c+2}\int\_{0}\&{1}dk~k\&{\c+2}\deriv{\o}{k}f(\o). \label{Avi}\EE
In the second integral--call it $\hat J$--we change variables to $\o$,
noting that, from the definition of $\l$ in eq.(\ref{Ao}) we have
\BE k\&{2}={\xi\&{2}\over p\&{2}+\o}+{\n\&{2}\over 1+\o}+{\z\&{2}\over
q\&{2}+\o}.  \label{Avii}\EE
Here $\xi\equiv x/Y,~\n\equiv y/Y,~\z\equiv z/Y$. At $k=0$
 $\o(0,\vec R)=\infty$.
We then get
\BE \hat J=-\int\_{\o(1,\vec R)}\&{\infty}d\o~f(\o)
\left({\xi\&{2}\over p\&{2}+\o}+{\n\&{2}\over 1+\o}+{\z\&{2}\over
q\&{2}+\o}\right)\&{{\c+2\over 2}}.  \label{Aviii}\EE
It is useful to change variables to $t\equiv 1/(1+\o)$, and
defining $\tau\equiv 1-p\&{2}$, $\k\equiv 1-q\&{2}$ we can write
\BE \hat J=-\int\_{0}\&{W}{t\&{(\c+1)/2}
[\xi\&{2}(1-\k t)+\n\&{2}(1-\k t)
(1-\tau t)+\z\&{2}(1-\tau t)]\&{(\c+2)/2}\over
[(1-\k t)(1-\tau t)]\&{(\c+3)/2}}~dt,  \label{Axax}\EE
where $W\equiv1/[1+\o(1,\vec R)]$.
 For $\vec R$ on the surface of the cutoff ellipsoid
$\o(1,\vec R)$ vanishes, so the upper integration boundary is 1.
\par
To obtain the acceleration at position $\vec R$
 inside the ellipsoid we make use of the fact
that the layers outside $\vec R$ do not contribute. Thus we can use the
above expression for the potential outside an ellipsoid that is cutoff
beyond that through $\vec R$, then take the $\vec R$ gradient,
and calculate the integrals at the boundary. When taking the $\vec R$
gradient, we note that the two contributions from differentiating the
boundaries of the integrals cancel, and we are left with the
following expression for the acceleration {\it inside} the cutoff
ellipsoid
\BE \vec g=-{2\pi AG\over \c+2}\grad [R\&{\c+2}I(\hat x,\hat y,\hat z)],
\label{Aviiia}\EE
with $I$ a function of the polar angles, but not of $R$.
\BE I=\int\_{0}\&{1}{t\&{(\c+1)/2}[\hat x\&{2}(1-\k t)+\hat y
\&{2}(1-\k t)(1-\tau t)+\hat z\&{2}(1-\tau t)]\&{(\c+2)/2}\over
[(1-\k t)(1-\tau t)]\&{(\c+3)/2}}~dt,  \label{Ax}\EE
where $\hat x=\sin\theta\cos\phi,~\hat y=\sin\theta\sin\phi$, and
$\hat z=\cos\theta$ are the angular factors of the coordinates $\vec R$.
This rather restricted form of the field follows from the same
 dimensional arguments we applied in sections 1 and 4.
The interior potential is then ($\c\not=-2$)
\BE \varphi=-{2\pi AG\over \c+2}R\&{\c+2}I(\hat x,\hat y,\hat z)+c.
\label{Aviccc}\EE
By comparison with the exterior potential at the boundary, where $W=1$
we get
 \BE c=-{2\pi AG\ro\&{\c+2}\over \c+2}\int\_{0}\&{\infty}f(t)~dt=
-{4\pi AG\ro\&{\c+2}\over \c+2}F\_{1}(1/2,1/2,1/2,3/2;\k,\tau).
  \label{gamza}\EE
\par
To get our results in the midplane of a
 disk we take $\theta=\pi/2$, and go to the limit
$q=0~(\k=1)$. All our expressions for the interior
 acceleration then follow.
Our expression for the potential outside an axisymmetric disk
eqs.(\ref{ti})(\ref{ssiii}) is obtained by putting $\tau=0$ and
 $Y=\ro$ in eqs.(\ref{Aiil})(\ref{Avi})(\ref{Axax}), and then writing
 the resulting integrals in terms of hypergeometric functions.
\par
Regarding the field at an arbitrary position outside the ellipsoid,
we note that in eq.(\ref{Axax}) the term in square brackets in
 the numerator of the integrand may be
written in the form $(R/\ro)\&{2}(1-bt)(1-ct)$.
 Thus, by changing variables to
$t=sW$ we can write
\BE \hat J=-B\int\_{0}\&{1}s\&{(\c+1)/2}[(1-bWs)(1-cWs)]\&{(\c+2)/2}
[(1-\k Ws)(1-\tau Ws)]\&{-(\c+3)/2}~ds,  \label{njnb}\EE
with $B=(R/\ro)\&{2\a}W\&{\a+1/2}$.
This can be written in terms of the Lauricella function of type $D$,
which is a generalization of the hypergeometric function to many
 variables (see \cite{ext76} 2.3.6). The first integral in eq.(\ref{Avi})
 can also be brought to a simpler form, and we finally get for the
 potential outside the cutoff, power-law ellipsoid
\BEAL \varphi(\vec R)&=&{-MG\over \ro}{2\a+1\over 4\a}\{2W\&{1/2}
  F\_{1}(1/2,1/2,1/2,3/2;\k W,\tau W)-\nonumber\\
 &&-\left({R\over\ro}\right)\&{2\a}W\&{\a+1/2}(\a+1/2)\&{-1}\times
\nonumber\\&&   \times F\&{(4)}\_{D}(\a+1/2,-\a,-\a,\a+1/2,\a+1/2;
\a+3/2;bW,cW,\k W,\tau W)\}.  \label{jkg}\EEAL
We remind the reader that $W=(1+\o)\&{-1}$, where $\o$ is the
positive solution of $(x/\ro)\&{2}/(p\&{2}+\o)+(y/\ro)\&{2}/(1+\o)
+(z/\ro)\&{2}/(q\&{2}+\o)=1$.
\par
The exterior potential may be written in terms of hypergeometric
functions of three variables
 when we are in any of the symmetry planes of the ellipsoid.


\begin{thebibliography}
\bibitem[Binney and Tremaine 1987]{BT87}\reference
Binney, J., and Tremaine, S., 1987, {\it Galactic Dynamics}, Princeton
University Press, Princeton.
\bibitem[Evans 1994]{eva94}\reference
Evans, N.W., 1994, MNRAS, 267, 333
\bibitem[Evans and de Zeeuw 1992]{EZ92}\reference
Evans, N.W., and de Zeeuw, P.T. 1992, MNRAS, 257, 152
\bibitem[Exton 1976]{ext76}\reference
Exton, H., 1976, {\it Multiple
 hypergeometric functions and applications}, Ellis Horwood  Ltd.
\bibitem[Gradshteyn and Ryzhik 1980]{GR80}\reference
Gradshteyn, I.S., and Ryzhik, I.M., 1980, {\it Tables of integrals
series and products}, Academic Press.
\bibitem[Kalnajs 1972]{kal72}\reference
Kalnajs, J.A., 1972, ApJ, 175, 63
\bibitem[Magnus and Oberhettinger 1949]{MO49}\reference
Magnus, W., and Oberhettinger, F., 1949, {\it Formulas and theorems
for the functions of mathematical physics}, Chelsea.
\bibitem[McMillan 1958]{mac58}\reference
McMillan, W.D., 1958, {\it The theory of the potential}, Dover.
\bibitem[Mestel 1963]{mes63}\reference
Mestel, L., 1963, MNRAS, 126, 553
\bibitem[Milgrom 1989]{mil89}\reference
Milgrom, M., 1989, ApJ, 338, 121
\bibitem[Qian 1992]{qia92}\reference
Qian, E., 1992, MNRAS, 257, 581
\end{thebibliography}
\end{document}